# A New Explanation of the Mechanism of Hadley Circulation

Wei Huang [1]

[1] Independent Researcher

Correspondence Email: huangwei5934@gmail.com, wei.huang@colorado.edu

ORCID: https://orcid.org/0000-0002-1594-8529

A New Explanation of the Mechanism of Hadley Circulation

The Hadley circulation (or Hadley cell) is traditionally described as a large-scale atmospheric circulation phenomenon driven by differential heating of the Earth's surface: warm, moist air rises near the equator, diverges poleward in the upper troposphere, and subsides in the subtropics. In this article, the mechanism of the Hadley circulation is revisited and a new model is provided to explain its mechanism. The new model is based on a form of the atmospheric dynamic equation which substitutes pressure with temperature and density; thereby categorizing weather systems into thermal and dynamic systems. Such classification is useful for explaining large-scale weather systems such as the Hadley cell. The proposed explanation for the mechanism of the Hadley circulation argues that subtropical highs are the driving force of the Hadley cell, rather than the conventionally-believed ITCZ (Intertropical Convergence Zone). To support our theory, we analyze the atmospheric air density flux divergence with the results from the Community Earth System Model (CESM) and derive a new continuity equation by adding source/sink terms, in which evaporation serves as the air-mass source, and precipitation (condensation) as the air-mass sink. Results found that the equatorial easterlies could be linked to the solar diurnal cycle, demonstrating that the trade wind can be generated by the solar diurnal cycle, especially in the spring and fall seasons, as well as from the equatorial branch of the subtropical high.

**Key Points**

A new primitive equation governed by temperature and density is derived. Thermal and dynamic systems can be defined by mathematical formula.

The equatorial easterly, or trade wind, results from the solar diurnal cycle due to atmospheric vicinity.

A new continuity equation is proposed to include source/sink terms, where evaporation as air-mass producer, precipitation as air-mass consumer.

**Plain Language Summary**

A new method is introduced which substitutes temperature and density into the atmospheric primitive equation replace pressure, so the atmospheric motion is governed by temperature and density. Under these conditions, if atmospheric flow is mainly controled by the temperature field (the temperature gradient value is higher than that of the density), the system can be named as a thermal system. One can see that subtropical highs, and polar cold lows belong to thermal systems. If atmospheric motion is dominated by the density field, the system is named as dynamic system. In this way, we can see that dynamic systems include ITCZ, tropical cyclones, mid-latitude cyclones, and polar cold highs. Based on the definition above, we demonstrate that ITCZ is not a thermal system, but dynamic system. Therefore, the new Hadley circulation can be explained as follows: solar radiation heats the subtropical region surface (mostly ocean), then longwave radiation from the surface heats the atmosphere to form warmer temperatures and produce anticyclonic flow. The equatorial branch then flows equator-wards, bring water vapor to generate warm moist air that accumulates at ITCZ, form clouds and generate convection to consume the water vapor. The equatorial easterly is generated from solar diurnal cycle.

**Introduction**

The model of Hadley circulation was first proposed by Hadley (1735), and its history that was adapted by the atmospheric community is described in the publication, Meteorology of History (Persson, 2006). According to this history, the Hadley cell is described as a large-scale atmospheric circulation phenomenon driven by differential heating of the Earth's surface: warm, moist air rises near the equator, diverges poleward into the upper troposphere, and subsides in the subtropics (Johanson and Fu, 2009). The rising branch of Hadley circulation is known as the Intertropical Convergence Zone (ITCZ), which lies in the equatorial trough, a permanent low-pressure feature where surface trade winds, laden with heat and moisture, converge to form a zone of increased convection, cloudiness, and precipitation (Waliser and Jiang, 2015). Figure 1 shows the three cells of Hadley circulation with further explanation from Wikipedia.

> Many years of observation and numerical models have proven the existence of Hadley cell; however, our current understanding and explanation regarding its formation mechanism remain challenged in terms of differing hypotheses, including the following.

The ITCZ, as the rising branch of Hadley circulation with its location at the thermal equator, should follow the maximum solar radiation zone moving with season, as the sun's zenith angle swings from 23.5° S in winter (northern hemisphere) to 23.5° N in summer (northern hemisphere); however, in reality, the ITCZ does not move as far as this zone indicates.

The high solar radiation zone is much broader than the ITCZ.

The ITCZ is thought to transport a large amount of energy upward and then poleward at the upper levels that is sufficient to provide energy for a broad range of subtropical regions, as well as to heat the air near the surface of subtropical region. However, with such a narrow ascending zone of the ITCZ and the relatively low air density at the upper levels, this phenomenon seems unlikely.

It is generally believed that the energy brought up by the ITCZ is from latent heat due to evaporation near the ITCZ. Consequently, the solar heat in the ITCZ and nearby regions cause the sea surface temperature (SST) to increase, which in turn enhances evaporation and cloud formation, eventually developing into convection and precipitation. However, that process is questionable in that the clouds tend to block the solar radiation from reaching the surface. Moreover, the temperature of rain itself is lower than that of the SST, which would cause SST to decrease, thus generating less water vapor and decreasing the formation of clouds, allowing more solar radiation to reach the surface. These described opposing processes would be expected to cause excessive fluctuation of clouds and precipitation; however, this is not consistent with the observed stable cloud masses in the ITCZ.

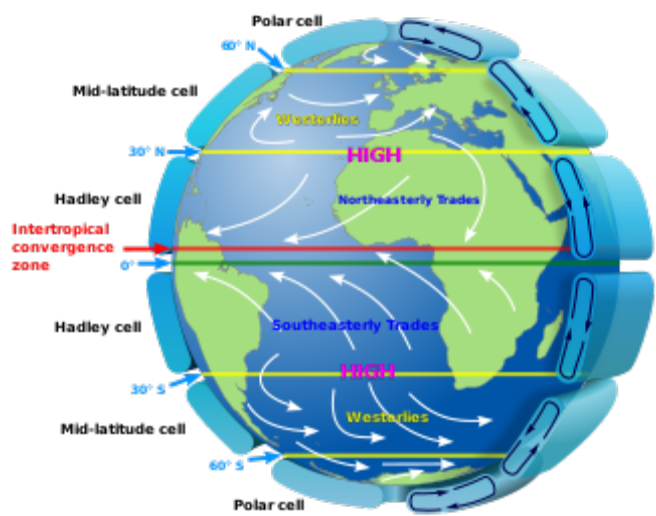

Image from: https://en.wikipedia.org/wiki/Hadley_cell

**Figure 1 Hadley cell**

As shown in Figure 1, the ITCZ is typically located near the northern hemisphere (especially during the northern hemisphere summer when the ITCZ is located around

10° N), thus air should be persistent across the equator. As the air crosses the equator from the south, the Coriolis force will cause air flow turn to the right-hand side, resulting in a southwesterly flow. However, no such southwesterly is observed on the southern side of the ITCZ. Or, at least, Fig. 1 does not illustrate such southwesterly. The traditional explanation of the Hadley circulation cannot explain the double ITCZ (Zhang, 2001), especially in the Spring or Fall season.

>In this article, a new explanation for the mechanism of the Hadley circulation has been provided. To do so, , the atmospheric dynamic equation is re-written by substituting pressure with temperature and density, which results in a clearer categorization of the differentiated weather systems and explains how large-scale weather systems are formed and maintained. Based on this new method, we propose a model of Hadley cell and ITCZ circulation that is different from the traditional model suggested by Hadley (1735).

One important phenomenon in the tropical region is the equatorial easterly also known as trade wind that is traditionally regarded as part of the Hadley cell. In this study, we provide a new mechanism which explains the trade wind in accordance with the solar diurnal cycle. The proposed new model of the Hadley circulation is verified by the Earth's evaporation and precipitation as produced by the Community Earth System Model (CESM version 1.2.2.1), thereby explaining their relation with the continuity equation, and their roles in modulating the Hadley cell.

**A new way to understand weather systems**

In order to facilitate our explanation of the weather systems in the atmosphere, the following dynamic equation will be re-written (Holton and Hakim, 2012), as done by Huang:

$$\frac{d\vec{V}}{dt} + 2\Omega \times \vec{V} = -\frac{1}{\rho}\nabla P + \vec{g}, \tag{1}$$

by substituting the state equation

$$P = \rho RT, \tag{2}$$

into the dynamic equation (1), and a new form of dynamic equation is obtained as follows:

$$\frac{d\vec{V}}{dt} + 2\Omega \times \vec{V} = -R\nabla T - RT\nabla \ln \rho + \vec{g}. \tag{3}$$

Considering the horizontal components of equation (3), the following horizontal dynamic equation is obtained:

$$\frac{d\vec{V_h}}{dt} + 2\Omega \times \vec{V_h} = -R\nabla_h T - RT\nabla_h \ln \rho, \tag{4}$$

where $\vec{V_h}$ is horizontal wind, and $\nabla_h$ is horizontal gradient. From the right-hand side of the equation (4), we can see that the effect of the pressure gradient is the sum of the temperature and density gradients. Based on Equation 4, the atmospheric motion can be in two modes:

Temperature and density gradient have the same sign (point to the same direction), which we can this as in Thermal-Dynamic mode
Temperature and density gradient have opposite sign (point to the opposite direction), which we name this as non-Thermal-Dynamic mode. In this paper we focus on the non-Thermal-Dynamic mode systems.

For non-Thermal-Dynamic systems, the temperature gradient and density gradients have opposite signs, i.e., either warm temperature (center) with low density (center), or cold temperature with high density; however, one gradient must have the same sign as the pressure gradient. Based on this, we can define the non-Thermal-Dynamic systems into two types as follows:

Thermal system: the temperature gradient has the same sign as the pressure gradient. In such system, temperature gradient serves as the major role.

Dynamic system: the density gradient has the same sign as the pressure gradient, where for such system, density plays the leading role.

From the definition above, we see that non-Thermal-Dynamic weather systems belong to one of four kinds as follows:

Warm (temperature center) high (pressure), but low density.
Cold (temperature center) low (pressure), but high density.
Cold (temperature center) high (pressure), but high density.
Warm (temperature center) low (pressure), but low density.

Kinds (i) and (ii) belong to the thermal system, as the high/low pressure is dominated by the temperature field; and kinds (iii) and (iv) are dynamic systems, as the density field controls the high/low pressure.

The above definition applies to all vertical levels. In this paper, we mainly focus on the surface level (or more strictly, the sea level), unless mentioned otherwise. According to the new definitions given above, typical weather systems can be categorized as:

Subtropical high: Warm center (likely with a low-density center); thermal system.

Polar vortex: Low temperature at its center (possibly higher density because it is located at higher latitudes); thermal system.

Winter polar (high-latitude) high: Cold center and high density; dynamic system.

Warm low: Tropical cyclone (hurricane, typhoon, or other tropical storm) and mid-latitude cyclone all belong to this category. From this point of view, the tropical cyclone and mid-latitude cyclone are identical storm types, yet occur at different latitudes.

Following these definitions, we can see that the thermal systems could be more closely related to energy budget (warm high, when the air has net energy gain; cold low, when the air has net energy loss), whereas dynamic systems could be more closely related to air density, in which the highs have higher density, and the lows have lower density. The atmosphere has a tendency to reduce the density difference, which is a process regarded as dynamics.

> Based on the above hypothesis, in the next section we are going to discuss the subtropical high and ITCZ, which are essential components of the Hadley circulation, as well as the mechanism of Hadley circulation.

**Mechanism of Hadley circulation**

A review of the processes related to solar radiation of Earth has first been presented before the issues listed in the introduction are addressed. The distribution of solar radiation at the top of the atmosphere is determined by Earth's sphericity and orbital parameters[13]:

$$\cos(\Theta) = \sin(\phi)\sin(\delta) + \cos(\phi)\cos(\delta)\cos(h), \quad (5)$$

where $\Theta$ is the solar zenith angle, $\phi$ is latitude [from -90° (south-pole) to 90° (north-pole)], $\delta$ is the solar declination [from -23.5° (winter solstice) to 23.5° (summer solstice)], and h is the hour (or relative longitude).

> By integrating Equation 5 over 24 h, and multiplying by the solar constant, the daily average solar radiation reaching the top of the atmosphere (known as clear sky radiation) can be obtained, which is shown in Fig. 2 for the following seasonal periods: spring/autumn equinox, summer solstice, winter solstice, between spring and summer, and between autumn and winter. The unit is the same as that of the solar constant, which is 1.361 kW/m².

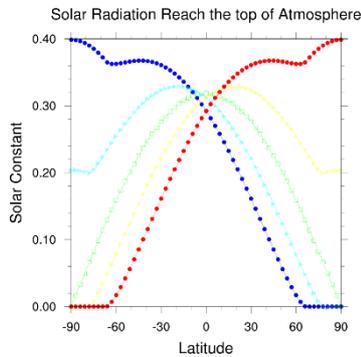

**Figure 2 Clear sky solar radiation at: spring/autumn equinox (green), summer solstice (red), and winter solstice (blue); and between spring equinox and summer solstice (yellow), and autumn equinox and winter solstice (cyan).**

In spring, the equator receives the highest level of solar radiation, but at the summer solstice, the warmest point is not directly under the sun near 23.5°, but at the higher latitudes around 40°. Another of the warmest points would be assumed to be at the pole; however, under realistic conditions the pole would not exhibit the maximum temperature because of the high reflectivity in that region. Although Fig. 2 shows that at the spring (or autumn) equinox, the equator receives more solar heating than in other regions, the ITCZ is not at the equator. Moreover, at the summer (or winter) solstice, the ITCZ is at only about 10°, which is far from the midpoint position of 23.5° and further still from the maximum radiation zone near 40°.

To study the distributions of solar radiation, long wave radiation, rain rate, sea level pressure (SLP), and other atmospheric fields related to Hadley circulation, we conducted global simulations using the CESM Version 1.2.2.1 (NE120_T12 grid). We ran CESM for several years; however, only the third-year results were used in this study (unless mentioned otherwise).

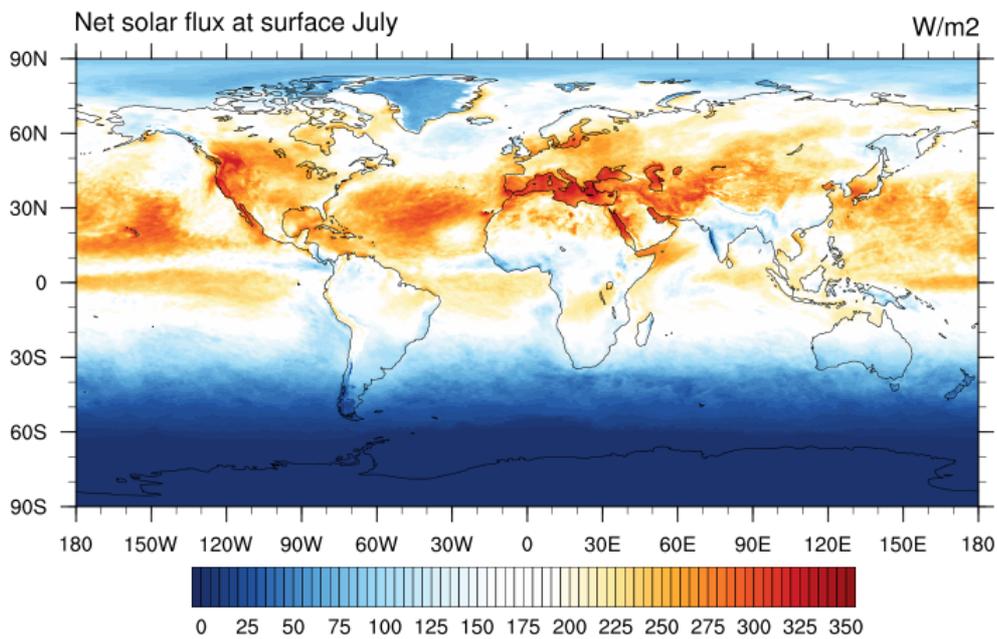

**Figure 3 Net solar flux at surface (w/m²) for July.**

Figure 3 shows the net solar flux at the surface in July from CESM. We see that the solar flux was zero below 60° S (i.e., the Antarctic region), which is consistent with the red line in Figure 2. Different from the red line on Figure 2 is that there was no large solar flux above 60° N, as the solar angle was smaller at high latitudes and the reflectivity was large. It is obvious that the zone of large solar flux was generally between 20–50° N, although the location of the maximum value varied in different regions; for example, the maximum value was located near 30° N at the Atlantic Ocean and central Pacific Ocean, and close to 40° N in the regions of the Mediterranean Sea and the mid-east. Between 5° and 10° N, a minimum solar flux band existed, from the west Pacific to continental Africa, which corresponded to the position of the ITCZ.

Figure 4 shows the rain rate of July from CESM. If ITCZ is defined by the rain band, we can see that the ITCZ corresponded to the region of the minimum solar flux band shown in Fig. 3. Further, a rain belt was present in the southern hemisphere; however, it resided only over the Indian and west Pacific Oceans. In comparing Figs. 3 and 4, we see that the ITCZ had less solar flux, which would be caused by the cloud-top reflection of solar radiation in the ITCZ. Clearly, the ITCZ was not located at the maximum solar flux zone, thus less solar flux would be reaching the surface at the location of the ITCZ; and therefore, the ITCZ would not be a direct response of solar radiation.

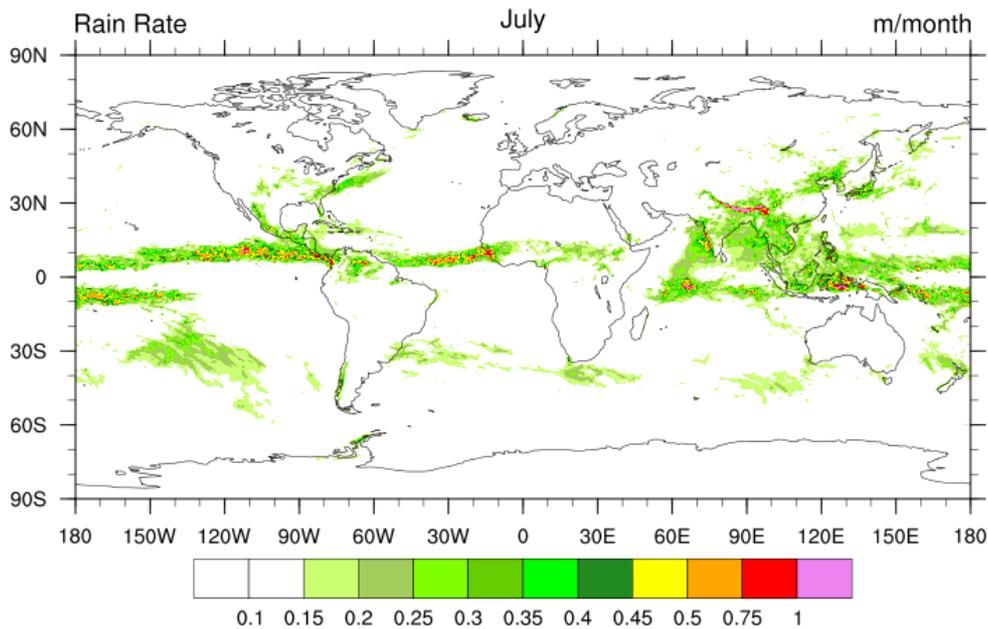

**Figure 4 Rain rate of July.**

It could be argued that in this scenario, although the ITCZ had less solar flux, it could receive more long wave radiation flux because clouds absorb more but emit less long wave radiation. Figure 5 shows the sum of the short (solar) and long wave radiation fluxes at the surface in July, where the positive and negative values indicate energy gain and loss at the surface, respectively. The maximum gain was 257.4 w/m$^2$ while the minimum loss was -85.7 w/m$^2$. Similar to Fig. 3, the area of the largest net radiation gain was around 30° N in the Atlantic Ocean and extended from 20° to 40° N in the Pacific Ocean. The minimum (white) zone was between 5° and 10° N, and extended from the west Pacific to Africa, where the ITCZ was located. Therefore, even considering both the solar and long wave radiation fluxes, the ITCZ still had less net radiation gain than its surrounding area.

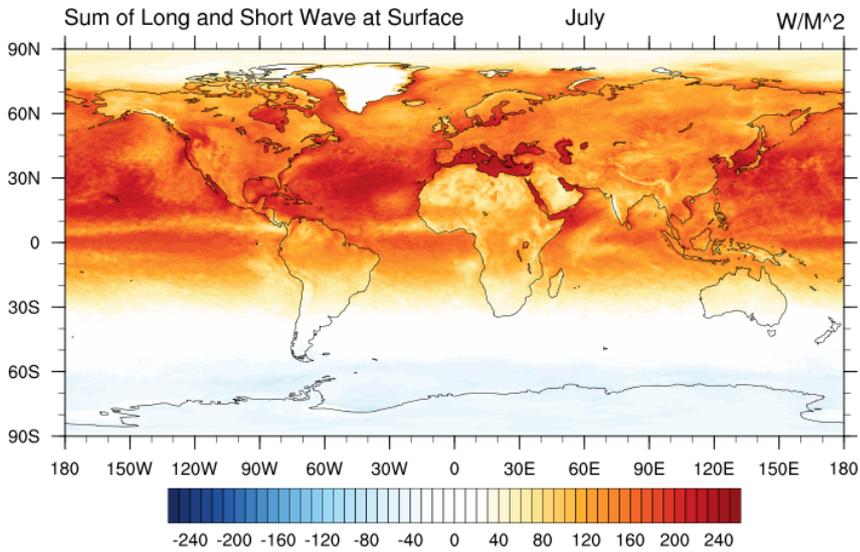

**Figure 5 Sum of short and long wave radiation (w/m²) for July.**

Figure 5 shows the sum of solar and long wave radiations in July. The maximum radiation gain was at 30° S and the minimum band was near 5° N, representing the ITCZ. The maximum gain was 276.7, and the maximum loss was -92.9.

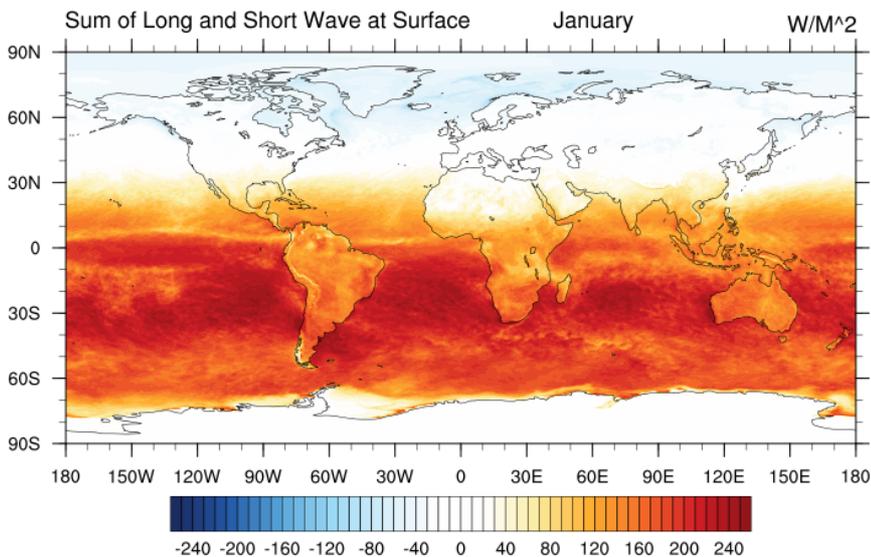

**Figure 6 Sum of short and long wave radiation (w/m²) for January.**

Figure 6 shows the sum of solar and long wave radiations for January. The maximum gain was located near 30°S and the ITCZ was located near 5°N. The maximum gain was 231.2, and the maximum loss was -84.4.

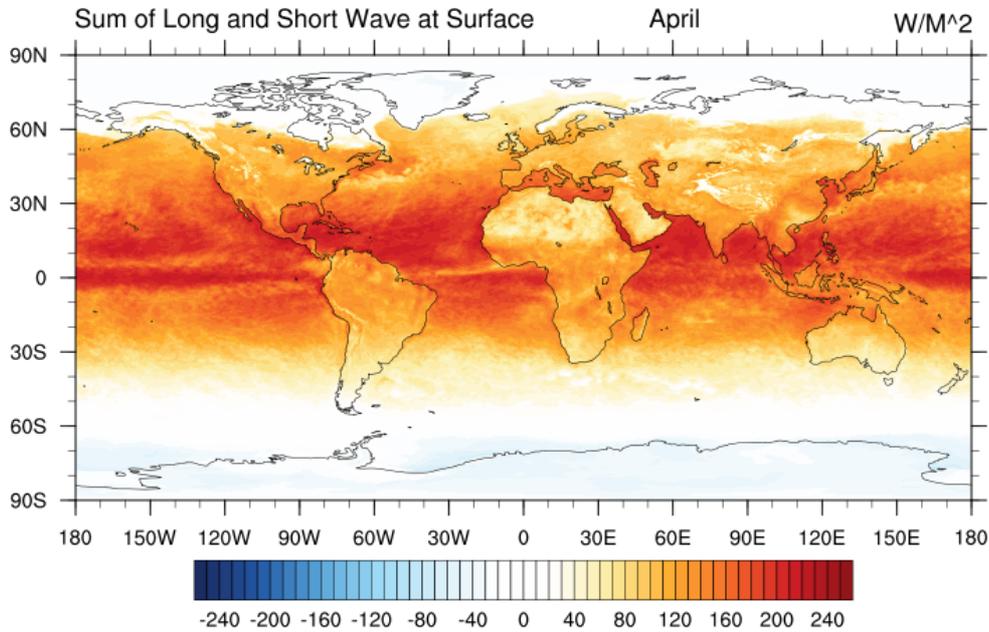

**Figure 7 Sum of short and long wave radiation (w/m²) for April.**

Figure 7 shows the sum of solar and long wave radiations for April. The maximum gain was located along the equator and the ITCZ was located near 5° N.

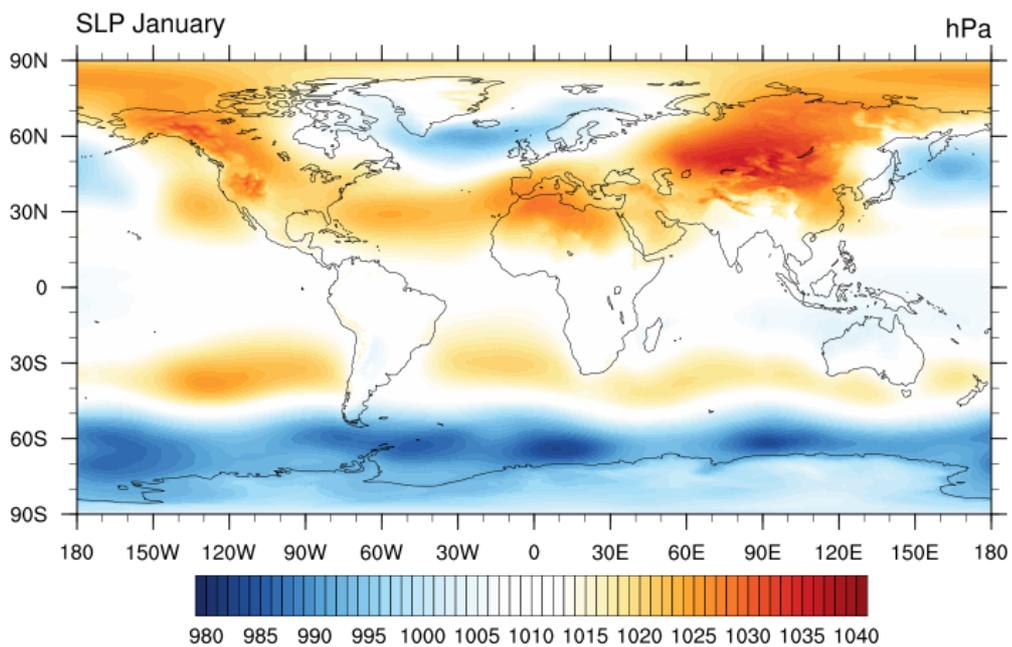

**Figure 8 Sea level pressure (SLP) for January.**

Figure 8 shows the SLP in January. Focusing on the high-pressure zone along 30° S, which is the area of the maximum net solar and long wave radiation zone, it is suggested that the net energy gain at 30° S produced higher pressure, rather than lower pressure. In July (figure omitted), the subtropical high in the northern hemisphere was located at higher latitudes than 30–40° N in the Atlantic Ocean and 45° N in the Pacific Ocean.

From Figs. 6 to 8, it can be seen that the net energy gain was much larger than the energy loss. A portion of this energy would be contributed to the internal energy increase by the following mechanism. As the surface absorbs the energy, its temperature would increase and enhance the water vapor flux, resulting in energy transmission into the air through the latent heating process.

Figure 9 shows the evaporation rate of July. Again, the evaporation rate at the location of the ITCZ (around 10°) was lower. In other words, the maximum evaporation region was not fully collocated with the maximum net radiation zone, which is reasonable because the warming of SST lags behind the net radiation due the time required for the ocean to warm up. One can see that the maximum evaporation zone was between 10° and 30° S in July, corresponding to the region where the warm SST formed during the southern hemisphere summer. Similarly, in January, the maximum evaporation zone was in the northern hemisphere at 5–25° N (figure is not shown).

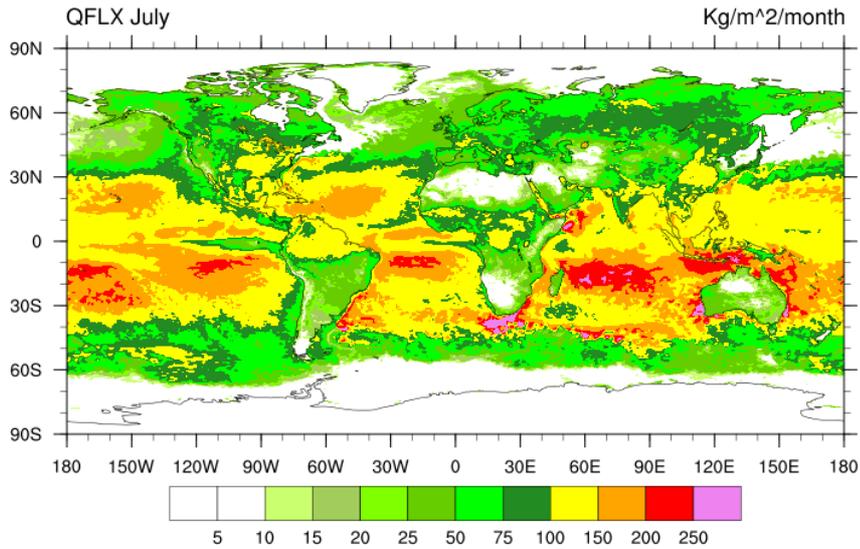

**Figure 9 Evaporation rate for July.**

Figure 10 shows the difference between January SST and the annual mean SST. The positive zone between 30–40° S was consistent with the energy gain as shown in Figure 6. Also apparent is the negative temperature difference between 30–60° N. Therefore, in January, the solar radiation was stronger in the southern hemisphere, which would be absorbed by the ocean and stored as internal energy, then released in the winter (northern) hemisphere. Such released energy in winter would provide the source to maintain a subtropical high and generate water vapor (in the winter hemisphere).

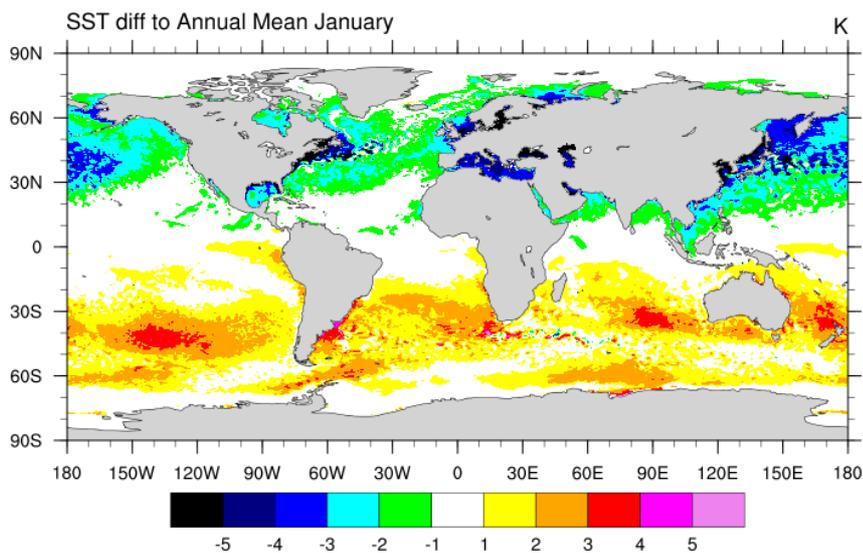

**Figure 10 Sea Surface Temperature (SST) Difference of January to the Annual Mean SST.**

Based on the above results, we can introduce a new conceptual model for the Hadley cell by assuming that the air is in a static state (i.e., no motion, no temperature, and no density gradient). When net solar radiation accumulates at the surface, the temperature in the higher radiation zone will increase, generating a temperature gradient (note that without a wind field to produce convergence/divergence, the density gradient would not be possible). Such a warm temperature center would eventually produce a divergent flow, and an anti-cyclone. As the divergent flow would cause a decrease in the air density, its gradient would balance the temperature gradient, creating a stable flow pattern. At this warm zone, with higher SST, the evaporation would be high and the air flowing out would contain more moisture. The moist air would flow equatorward and accumulate there; subsequently forming clouds, convection, and heavy rainfall. Because the subtropical high region would remain clear of cloud cover, the solar radiation could reach the ocean surface during the daytime, allowing the moist air to be generated continuously. Since the clouds and convection at the ITCZ would be supplied by moist air without interruption, the process would be stable. Since the latent heat release would be more efficient than the latent heat obtained from evaporation, the ITCZ could remain narrow. In the summer, even with the sun at 23.5° and having a longer daytime at higher latitudes, the maximum solar radiation zone would be at higher latitudes of 30–35°, which would be the location of the subtropical high region. It should be noted that moist air has a lower density than dry air at the same temperature; therefore, the ITCZ would become a lower-pressure region as moisture accumulates in that region.

It is worth mentioning that the poleward flow (on the high-latitude side of the subtropical high region) at the surface would be expected to form a mid-latitude cyclone, producing precipitation as it moved.

Certain issues which challenge this new explanation of the Hadley cell mechanism include the following:

High pressure at the subtropical region has a divergent flow, which would cause density loss; however, the descending motion at this region could not fully compensate for the air density loss, as the density flux at upper levels is much smaller than that at lower levels because of the density difference.

The convergent flow at the ITCZ would result in air-mass accumulation; however, the upper-level divergent flow pressure would not be sufficient to remove the air-mass accumulated, also due to the density difference between the upper and lower levels.

As the Coriolis force approaches zero near equator, the equatorward flow may not turn quickly enough to become easterly; therefore, the flow may continue on across the equator (from the summer hemisphere to the winter hemisphere).

In an attempt to address issues 1 and 2 above, consider that the mass loss in the subtropical high region is compensated by evaporation since the moisture enters the atmosphere and serves as an air-mass source, or air-mass producer. In contrast, rain at the ITCZ has an opposite effect. When condensation occurs, producing precipitation, the air loses moisture, and thus decreases its total mass. Therefor the precipitation in the ITCZ serves as an air-mass sink, or air-mass consumer.

To further examine the air-mass change at the subtropical high and ITCZ, let us review the continuity equation, as Huang did:

$$\frac{\partial \rho}{\partial t} + \nabla \cdot (\rho \vec{V}) = 0. \tag{6}$$

Eq. 5 can be re-written as:

$$\frac{\partial \rho}{\partial t} = -\nabla \cdot (\rho \vec{V}). \tag{7}$$

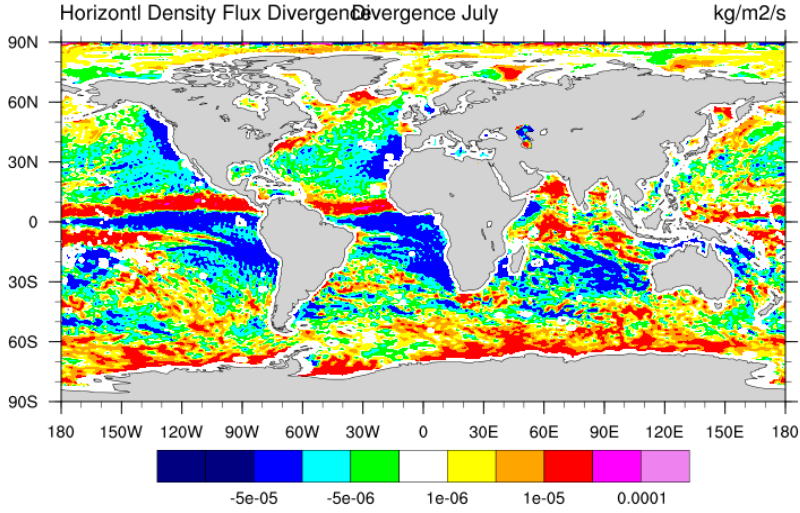

**Figure 11 Density flux divergence for July**

Figure 11 is the result of vertical integration of the right-hand side of Eq. 7 using CESM results. A narrow band of positive density (mass) tendency can be seen at the ITCZ, as well as a negative area at the subtropical region, which provides evidence for the reasoning above.

Because evaporation provides air-mass, it therefore serves as a density source, and such precipitation reduces air-mass which serves as a density sink, the continuity equation can be altered to a new form, as:

$$\frac{\partial \rho}{\partial t} + \nabla \cdot (\rho \vec{V}) = Source + Sink. \tag{8}$$

Huang (2016) pointed out the impacts of the solar diurnal cycle, from Eq. 4, and focus on the thermal term, but now for the daily change. In the daytime, the solar radiation is strong and the air temperature is higher than that at nighttime when there is no solar radiation and long wave radiation causes the air temperature to drop. Consequently the pressure is relatively higher in the daytime regions and lower in the nighttime regions, resulting in a temperature gradient between the two regions and causing the air to flow from the daytime region to the nighttime region. Given the viscosity of air and the rotation of the Earth, the heated region is always inclined westward and therefore the wind would be stronger on the west side of the daytime region compared to that on the east side, thus producing easterly winds. The easterly winds would block the equatorward flow from the subtropical region and prevent the low-level air flows from the summer hemisphere to the winter hemisphere, an essential mechanism for the ITCZ. This not only explains the obvious lack of cross-equator flow and westerly winds on the equator-side of the ITCZ, but also the reason whys ITCZ is located in the summer hemisphere.

One way to prove the discussion above is to do a simulation with an aqua planet. Based on our discussion above, we should see few things: a) the multiple yearly averaged circulation should have two ITCZs, or more strictly, if one makes an idealized experiment with a fixed Spring season, which the sun is always located above the equator, there should be two ITCZs, with equatorial easterly, and divergency near equator. Recent work (Wu etc., 2021) just approved these. 100 years zonally averaged zonal wind from Wu and her colleague's work on aqua planet simulation shows equatorial easterly, double ITCZs, with divergency near equator. One can further thinking, if solar radiation has no diurnal cycle, the equatorial easterly would not appear, and it should be quiet there. Instead, one would see weak westerly at the equatorial side of the ITCZs.

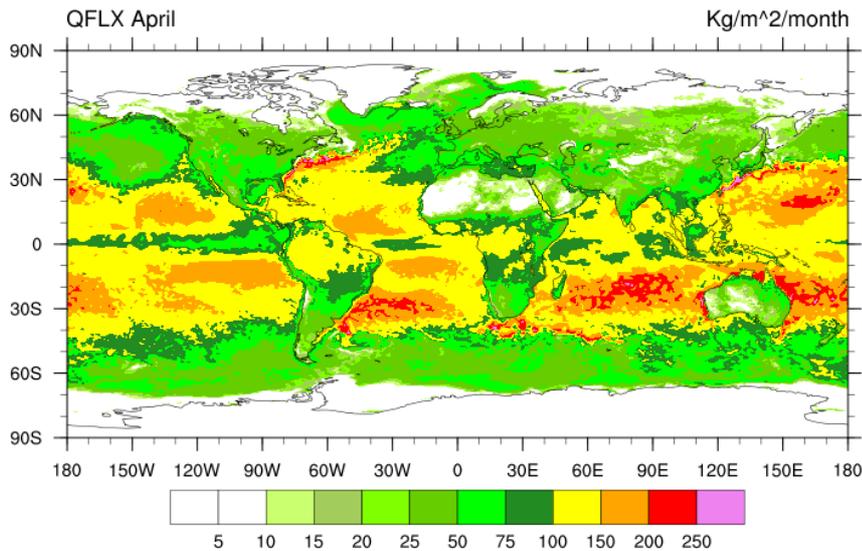

**Figure 12 Evaporation rate for April from CESM.**

In the Spring/Fall season, when the sun is above the equator, easterly winds generated by the diurnal cycle are stronger along the equator, causing ocean upwelling and easterly oceanic current along that region. The Coriolis force transforms the easterlies into south-easterlies in the northern hemisphere, and into north-easterlies in the southern hemisphere. When such easterlies also meet the easterlies from the subtropical high, two ITCZs are formed, a phenomenon revealed by observations of Zhang (2001) that showed double ITCZs in the Pacific Ocean during spring and fall. There is little evidence of double ITCZs in the Atlantic Ocean, which could be due to the size and shape of the ocean and the influence of the African continent. Figure 12 shows the evaporation rate in April and identifies that the minimum evaporation rate was located along the equator, likely due to the low SST from ocean upwelling, even in the presence of strong solar radiation at the equator during the Spring and Fall.

We now summarize the main components of the proposed new mechanism for the Hadley cell:

Solar radiation heats the surface and causes the surface temperature to rise, and at the same time the Earth releases longwave radiation to heat the air and cool itself down, which results in a net radiation gain in the summer hemisphere and net radiation loss in the winter hemisphere. In the summer hemisphere, the net radiation gain region is a broad zone, with its maximum extent at the 35–40° latitudes.

Once the warm surface has formed, the surface pressure increases to form the subtropical high that generates a surface divergent flow equatorward at lower latitudes (south of the high center) and poleward at the high latitudes. The divergent flow also causes descending motion and air-mass loss. The latter is then partially compensated for by water vapor transported from the surface.

If the surface is ocean, then the net solar radiation gain heats the water and produces water vapor, then the water vapor is carried both equatorward and poleward. The equatorward flow becomes north-easterlies, while the poleward flow becomes south-westerlies.

The diurnal cycle of solar radiation generates easterlies in the equatorial region. This is like stir the water with a stick in the pond, keeping stir the water in circle, will generate a circulation. Where when the sun heats the equatorial region in a circle, it produces a circulation around the earth, which is the equatorial easterly.

The easterlies block the equatorward flow, which carries moist air and causes the moist air accumulated to form the ITCZ. The convergence at ITCZ causes an air-mass gain that is consumed by producing precipitation.

Condensation processes at the upper levels of the ITCZ release latent heat, warming the upper-level air, and forming upper-level high that produces divergent flow going further equatorward or poleward. The poleward flow becomes westerlies due to the Coriolis effects, while the air descends at the top of the subtropical high region.

**Further work**

Based on the new way to understand the weather systems represented by Eqs (1–4), such as the Hadley cell, we can apply a similar idea to the understanding of other meteorological systems such as polar vortex. In the modern explanation of meridional circulation, it is said that the polar region forms in the descending part of the polar cell. However, we believe that the low in the polar region is caused by longwave radiation, especially in the winter season. The analysis will be presented in a different paper.

**Conclusion**

We listed the problematic issues in the current explanation of the Hadley circulation and proposed a new conceptual model that has been shown to mitigate some of the conflicting issues. We first introduced a new form of the dynamic equation, showing that wind is controlled by the combined effects of temperature and density gradient. Based on that equation, we defined a new method to categorize the weather systems into two types: 1) Thermal systems, in which the air motion is influenced more by the temperature gradient term; and 2) dynamic systems, in which the air motion is dominated by the density gradient term. With this classification method, we then conducted an analysis to argue that the subtropical high is a thermal system forced by the net gain between the solar and long wave radiation. Conversely, we infer that the ITCZ is a dynamic system, for which the density gradient plays a more important role in the ITCZ formation and maintenance. Whereas the summer subtropical high is a direct thermal response to net radiation, the winter subtropical high is an indirect thermal response because in winter the heat originates from the release of stored internal energy, mainly in the oceanic regions.

Using CESM results, we diagnosed the density flux divergence and source/sink term to the continuity equation, providing evidence that evaporation in the subtropical region serves as the air-mass source and precipitation as the air-mass sink. Further, it was inferred that solar diurnal cycles produce easterlies at the tropics. The easterlies generated from the solar diurnal cycle at the equator are stronger, which helps to explain the observed double ITCZs.

Here, a new picture of Hadley circulation is presented: The subtropical high is the direct response to solar radiation, which warms the temperature at this region, produces divergent flow, and provides moist air to compensate for the air loss from the divergent flow. The equatorward flow then brings moisture to lower latitudes, together with the easterlies at the tropical region to form the ITCZ. The ITCZ precipitation serves as a way to reduce the air accumulation at lower levels.